\documentclass[11pt]{article}
\usepackage{amssymb,cite,graphics,a4}
\newlength{\indentedwidth}
\newdimen\mathindent
\indentedwidth=\mathindent

\DeclareMathAlphabet{\mathpzc}{OT1}{pzc}{m}{it}  
\begin{document}
\vskip 0.5cm
\begin{center}
{\Large \bf Integrability of the russian doll BCS model} 
\end{center}
\vskip 0.8cm
\centerline{Clare Dunning%
\footnote{\tt tcd@maths.uq.edu.au} 
and Jon Links%
\footnote{\tt jrl@maths.uq.edu.au}}
\vskip 0.9cm
\centerline{\sl\small 
 Centre for Mathematical Physics, School of Physical Sciences, }
\centerline{\sl\small The University of Queensland, Brisbane 4072,
 Australia.} 
\vskip 0.9cm
\begin{abstract}
\vskip0.15cm
\noindent
We show that integrability of the BCS model extends beyond
Richardson's 
model (where all Cooper pair scatterings have equal coupling) to 
that of the russian doll BCS model  for which the couplings have a 
particular phase dependence that breaks time-reversal symmetry.
This model is shown to be integrable using the quantum inverse scattering
method, and the exact solution 
is obtained by means of the algebraic Bethe ansatz. The inverse problem
of expressing local operators in terms of the global operators
of the monodromy matrix is solved. This result  
is used to find a determinant formulation of a 
correlation function for fluctuations in the Cooper pair occupation
numbers. These results are used to undertake exact numerical analysis for 
small systems at half-filling.

\end{abstract}

\setcounter{footnote}{0}
\def\thefootnote{\fnsymbol{footnote}}

\def\d{\dagger}
\def\e{\varepsilon}
\def\l{\left}
\def\r{\right} 
\def\t{\tilde{t}}
\def\ve{\varepsilon}
\def\beq{\begin{equation}}
\def\eeq{\end{equation}}
\def\bea{\begin{eqnarray}}
\def\eea{\end{eqnarray}}

\def\le{\langle}
\def\re{\rangle}
\def\lt{\left}
\def\rt{\right}
\def\o{\omega}
\def\nn{\nonumber} 
\def\L{\mathcal{L}}
\def\g{{\mathfrak{g}}}
\def\La{\Lambda}
\def\LLa{\tilde{\Lambda}}
\def\eps{\varepsilon}
\newcommand\eq{\begin{equation}}
\newcommand\en{\end{equation}}
\newcommand\alp{\alpha}
\newenvironment{fig}{\linespread{1.0} \begin{figure}}{\end{figure}%
\linespread{1.3}}
\newcommand{\fl}{\hspace*{-\mathindent}}
\newcommand\phup{^{\phantom p}}

\section{Introduction}

Experimental work conducted by  Ralph, Black and Tinkham 
\cite{brt,rbt} was successful in determining the discrete energy 
spectra of aluminium 
nanograins with a fixed number of electrons.  Their results 
surprisingly showed  
 remnants of superconducting behaviour. Significant differences between the
spectra of 
grains with even or odd number of electrons suggested that
strong pairing correlations characteristic of superconductivity were
still present, and quantum fluctuations  
significant.  The 
Bardeen, Cooper and
Schrieffer (BCS) model of superconductivity  \cite{bcs} can be applied
in this situation, but the usual mean-field approach is not
suitable for two reasons: firstly 
 the number of electrons must be fixed, and secondly the strong quantum
fluctuations 
destroy the validity of the assumption that operators may be
averaged.  

Fortunately predictions from 
Richardson's model,  
for which 
all scatterings between pairs are assumed to have equal
coupling, 
give an excellent match with 
the experimental results discussed above \cite{vdr}. 
In particular, this model is exactly solvable. 
 The  solution was originally 
obtained and analysed in the 1960s by Richardson and Sherman in the
context of nuclear physics \cite{richardson,rs}. 
The 
 model has also been shown to be integrable \cite{crs}, and by recasting
 this result in the context of the quantum inverse scattering method,  
the solution was subsequently obtained by  means of the algebraic
Bethe ansatz method \cite{zlmg,vp}.  
The latter approach has the advantage that 
it is possible  to derive exact expressions for form factors and
correlation functions \cite{zlmg,ao,lzmg}.  These considerations 
have 
already been used to produce several
generalisations of Richardson's model \cite{ado1,ado2,des}, 
and to study other pairing Hamiltonians \cite{lzgm,gflz}.  
Reviews on these topics can be found in \cite{vdr,lzmg,dps}. Exact
studies of these BCS models are also important for studying 
pairing correlations in systems of ultracold dilute Fermi gases \cite{rgj}. 

Here we study the russian doll BCS model, a one-parameter generalisation  of 
Richardson's model.    
It differs from Richardson's model 
by the inclusion of phases 
(indexed by a parameter
$\alpha$) in the
pair scattering couplings, which break time-reversal symmetry.  
Richardson's model is reproduced in the limit as the parameter
$\alpha$ is taken to zero.  The terminology `russian doll'  
refers to the russian craft of making nested wooden dolls, known as 
{\it matryoshka}. Inside each doll one finds a smaller version of the
doll. A russian doll renormalisation group flow is 
one which displays
a cyclic nature rather than flowing to a fixed point. The russian doll
BCS model was proposed in    
\cite{lrs} as an example of a 
many-body system exhibiting a russian doll renormalisation group flow.
In \S 2, the model is shown to be
integrable via the quantum inverse 
scattering method and its exact spectrum is obtained via
the algebraic Bethe ansatz method. Our derivation shows that the
Hamiltonian is embedded in the transfer matrix as the
second-order term in the expansion about an infinite spectral parameter. 
This construction is in contrast to the proof of integrability for 
Richardson's model given in \cite{lzmg} in that the quasi-classical limit is
not taken. In this sense there is no connection between the russian doll
BCS model and
Gaudin magnets (cf. \cite{dps}). 
The effect of the parameter $\alpha$ 
on the spectrum of the model is discussed in \S 3.   
In particular we will show that $\alpha$ influences the
degeneracies of the energy bands in the strong coupling limit. 
In \S 4 we solve the inverse problem. 
This result  is used in 
\S 5 to obtain an exact formula for a superconducting pairing
correlation function. Numerical analysis of our 
results show that the ground state of the
model is qualitatively independent of $\alpha$ in terms of the
fluctuations in Cooper pair number for the single particle energy
levels. Increasing $\alpha$ does however lead to a quantitative 
suppression of the
fluctuations across all levels. 
Our conclusions are given in  
\S 6.

\section{Integrability of the russian doll BCS model and its exact solution}

The physical properties of a metallic nanograin with pairing
interactions are described by the
BCS Hamiltonian \cite{vdr}
\beq H_{\rm{BCS}}=\sum_{j=1}^{\L}\e_jn_j
-\sum_{j,k=1}^{\L}g_{jk}\phup 
c_{k+}^{\d}c_{k-}^{\d}c_{j-}\phup c_{j+}\phup. \label{bcs_eqn} 
\eeq
Here, $j{=}1,\dots,{\L}$ labels a shell of doubly degenerate single
particle
energy levels with energies $\e_j$, and $n_j=c^\dagger_{j+}c_{j+}\phup
 + \ c^\dagger_{j-}c_{j-}\phup $ is the
fermion number operator for
level $j$. The operators $c_{j\pm}\phup ,\,c^{\d}_{j\pm}$ are the annihilation
and creation operators for the fermions at level $j$. The labels $\pm$
refer
to pairs of time-reversed states.

An important aspect of the Hamiltonian (\ref{bcs_eqn})
is the blocking
effect. For any unpaired fermion at level $j$ the action of
the pairing interaction is zero since only paired fermions are
scattered. This means that the Hilbert space can be decoupled into
a product of paired and unpaired fermion states in which the
action of the Hamiltonian on the space for the unpaired fermions is
automatically diagonal in the natural basis.
In view of the blocking effect, it is convenient to introduce
hard-core boson operators
$b_j\phup =c_{j-}\phup c_{j+}\phup\, , b^{\d}_{j}=c^{\d}_{j+}c^{\d}_{j-}$
which satisfy 
the relations
$$ 
(b^{\d}_j)^2=0, \quad\quad [b_j\phup
,\,b_k^{\d}]=\delta_{jk}\phup (1-2b^{\d}_jb_j\phup )
,\quad\quad [b_j\phup ,\,b_k\phup ]=[b^{\d}_j,\,b^{\d}_k]=0 $$ 
on the space excluding single particle states. We also set
$N_j\phup =b^\dagger_j b_j\phup $ to be the Cooper pair number operator for the
$j$th level.
In this setting the hard-core boson operators realise the
$su(2)$ algebra in the pseudo-spin representation,
through the identification
\beq S^-=b , \quad  \quad 
S^+=b^{\dagger}, \quad  \quad
S^z=\frac{1}{2}\left(2N\phup -I\right). \label{psr} \eeq
The pseudo-spin operators satisfy the following commutation relations: 
\beq [S^z, S^{\pm}] = \pm S^\pm, ~~~~[S^+,S^-] =2 S^z.\label{su2} \eeq

It is now well known that the Hamiltonian (\ref{bcs_eqn}) can be solved
exactly in the case when $g_{jk}=g$ is constant for all $1\leq j,k \leq \L$.
This is referred to  as Richardson's model. The solution was obtained
in  \cite{richardson}. It was subsequently shown that the model is
also integrable through an explicit construction 
for a  set of conserved operators
\cite{crs}. Both of these results can be obtained from the
formulation of the model through the quantum inverse scattering
method \cite{zlmg,vp}. Here we will adopt this approach
to establish that there exists a more
general manifold of 
integrability. 
We will show that the
following Hamiltonian
\beq 
H_{\rm{RD}}=  2\sum_{j=1}^{\L} \e_j N_j- g \sum_{j<k}^{\L}
(e^{i\alpha}b_k^\dagger b_j\phup 
+e^{-i\alpha}b_j^\dagger b_k\phup ) \label{int} \eeq 
is integrable for arbitrary values of $\alpha$. 
This model includes Richardson's model in the limit
$\alpha\rightarrow 0$, up to the addition of a multiple of the $u(1)$ operator
$N=\sum_{j=1}^{\L} N_j$ for the total number of Cooper pairs. It
coincides with the russian doll BCS model
  recently studied in \cite{lrs}.

To establish the integrability of (\ref{int}), 
we begin with the $R$-matrix
$R(u)$ that satisfies the Yang--Baxter equation 
\eq
R_{12}(u-v)R_{13}(u)R_{23}(v)
=R_{23}(v)R_{13}(u)R_{12}(u-v),  \en
and  has the form \cite{mcguire,yang}  
\bea
R(u)& = & \frac{1}{u+i\eta}({uI\otimes I+i\eta P})\nn \\ [3pt]
&=& 
{\renewcommand{\arraystretch}{1.1} 
\left ( \begin {array} {cccc}
1&0&0&0\\
0&b(u)&c(u)&0\\
0&c(u)&b(u)&0\\
0&0&0&1\\
\end {array} \right )
},
\label{rm} \eea
where $b(u)=u/(u+i\eta)$,   
$c(u)=\eta/(u+i\eta)$ and $\eta$ is taken to be an arbitrary real parameter.
Above, $P$ is the permutation operator. It satisfies 
$$ P(x\otimes y)=y\otimes x $$  
for all vectors $x$ and $y$. 
An important property of the $R$-matrix that will be exploited later
is that it satisfies the
unitarity condition
\beq R(u)R(-u)= I \otimes I.  \label{unitarity} \eeq  

The $R$-matrix is $gl(2)$-invariant in that 
\beq [R(u),\,\g\otimes \g]=0 \label{gl2} \eeq 
where $\g$ is any $2\times 2$ matrix. 
We introduce the Lax operator $L(u)$ 
in terms of the $su(2)$ Lie algebra 
 with generators $S^z$ and $S^{\pm}$ \cite{korepin,faddeev},
\beq 
 L(u) = \frac{1}{u+i\eta}\Biggl ( 
{
\renewcommand{\arraystretch}{1.1}
\begin {array}{cc}
 uI+{i\eta} (S^z+I/2) &{i\eta} S^-\\
 {i\eta} S^+& uI+{i\eta} (I/2-S^z)\\
 \end {array} 
\Biggr )},
\label{ls} 
\eeq  
 subject to the commutation relations (\ref{su2}).  
Then the following holds: 
\beq R_{12}(u-v)L_1(u)L_2(v)=L_2(v)L_1(u)R_{12}(u-v). 
\label{rll} \eeq 
When the $su(2)$ algebra
takes the spin $1/2$ representation (\ref{psr})
the resulting $L$-operator is equivalent  
to that given by the $R$-matrix (\ref{rm}).  

We take 
$\g=\exp(-i\alpha {\sigma})$ with
$\sigma={\rm diag}(1,\,-1)$ and use  
(\ref{ls})
to  construct the monodromy matrix
\bea T(u)&=&\g_0L_{0{\L}}(u-\e_{\L})\cdots
L_{02}(u-\e_2)L_{01}(u-\e_1) \nn \\ [2pt]  
&=&\pmatrix{A(u)& B(u) \cr C(u) & D(u)}. \nn \eea 
Above, 0 labels the auxiliary space,
while the tensor components of the physical space are labelled
$1,\dots,\L$.
Thus the entries $A(u),\,B(u),\,C(u),\,D(u)$ of $T(u)$ are 
elements of the $\L$-fold tensor algebra of $su(2)$.
As a consequence of (\ref{gl2}) and (\ref{rll}) we have the relation  
\beq R_{12}(u-v)T_1(u)T_2(v)=T_2(v)T_1(u)R_{12}(u-v).\label{rtt} \eeq
We next define the transfer matrix 
\bea t(u)&=&{\rm tr}_0 T(u) \nn \\
&=&A(u)+D(u). \nn \eea 
Here, 
${\rm tr}_0$ denotes the trace
taken over the auxiliary space. 
 It can be shown 
using (\ref{rtt}) 
that the transfer matrix $t(u)$ generates  a
one-parameter family of commuting operators, viz. 
$$ [t(u),\,t(v)]=0~~~~~\forall\,u,v\in \mathbb{C}. $$ 
Next we specialise the representation of the $su(2)$ algebra to that given
by (\ref{psr}). 
Following the method of the algebraic Bethe ansatz (see for example
\cite{lzmg}) we can deduce that the transfer matrix eigenvalues are
$$ \La(u)=\exp(-i\alpha)
\prod_{k=1}^\L\frac{u-\ve_k}{u-\ve_k+i\eta}
\prod_{j=1}^N\frac{u-v_j+3i\eta/2}{u-v_j+i\eta/2} 
+\exp(i\alpha)
\prod_{j=1}^N\frac{u-v_j-i\eta/2}{u-v_j+i\eta/2} $$ 
subject to the constraints of the Bethe ansatz equations
\beq
\exp(-2i\alpha)\prod_{k=1}^\L\frac{v_i-\ve_k-i\eta/2}{v_i-\ve_k+i\eta/2}
=\prod_{j\neq i}^N\frac{v_i-v_j-i\eta}{v_i-v_j+i\eta}. \label{bae} \eeq

It is convenient to define an equivalent transfer matrix 
$$\t(u)=\left(\prod_{k=1}^\L\frac{u-\ve_k+i\eta}{u-\ve_k+i\eta/2}\right)
t(u), $$ 
which has eigenvalues 
$$ \LLa(u)= \tilde{a}(u)
\prod_{j=1}^N\frac{u-v_j+3i\eta/2}{u-v_j+i\eta/2} 
+\tilde{d}(u) 
\prod_{j=1}^N\frac{u-v_j-i\eta/2}{u-v_j+i\eta/2} $$   
where
\bea \tilde{a}(u)&=&
\exp(-i\alpha)\prod_{k=1}^\L\frac{u-\ve_k}{u-\ve_k+i\eta/2} 
\label{ta} \\
\tilde{d}(u)&=&\exp(i\alpha)\prod_{k=1}^\L\frac{u-\ve_k+i\eta}{u-\ve_k+i\eta/2}
\label{td} . \eea 
Analogously we define 
$$\tilde{A}(u)=\left(\prod_{k=1}^\L\frac{u-\ve_k+i\eta}{u-\ve_k+i\eta/2}\right)
A(u)$$ 
and similarly for $\tilde{B}(u),\,\tilde{C}(u)$ and $\tilde{D}(u)$.  

A series expansion for $\t(u)$ yields a set of mutually
commuting operators. 
Here we choose to expand about the point at
infinity: 
$$ \t(u)=\sum_{k=0}^{\infty} \t^{(k)} u^{-k} $$  
such that 
$$ [\t^{(k)},\,\t^{(l)}]=0 \quad \quad \forall\, k,\,l. $$ 
It is straightforward to deduce the leading terms:   
\bea 
&& \hspace{-1cm} \t(u)
\sim  2\cos(\alpha) \Biggl(I +\eta u^{-1} \tan(\alpha) (N-\L/2)
-\frac{i\eta^2 u^{-2}}{2} \tan(\alpha) (N-\L/2) \nn \\
&&~~~~~~~~~~~~~~~~  +\eta u^{-2} \tan(\alpha) \left[\sum_{i=1}^{\L}  
\e_i N_i
-\frac{1}{2}  \sum_{i=1}^{\L} \e_i
\right]\nn \\
&&~~~~~~~~~~~~~~~~-\eta^2 u^{-2} \biggl [\frac{1}{2}
(N{-}\L/2)^2-\frac{\L}{8}
+\frac{1}{2\cos(\alpha)} \sum_{j<k}^{\L}(e^{i\alpha}b_k^\dagger
b_j\phup 
+e^{-i\alpha}b_j^\dagger b_k\phup )\biggr ] \Biggr) \label{exp1}.
\eea
Expanding $\LLa(u)$ in powers of $u^{-1}$ we find
\bea \LLa(u)&\sim& 2\cos(\alpha)
\left(1 + \eta u^{-1} \tan(\alpha) (N-\L/2)
-\frac{i\eta^2 u^{-2}}{2} \tan(\alpha) (N-\L/2)\right. \nn \\
&&~~~~~~~~~~~~~~~~~~~~~+  \eta u^{-2}\tan(\alpha) \left[ \sum_{i=1}^N
v_i
-\frac{1}{2} \sum_{i=1}^\L \ve_i
\right]\nn \\
&&~~~~~~~~~~~~~~~~~~~~~\left.-\eta^2 u^{-2} \biggl [\frac{1}{2}(N-\L/2)^2
 -\frac{N}{2}  -\frac{\L}{8}\biggr] \right). \nn
 \eea
 Comparing this expression with the expansion (\ref{exp1})
 we immediately see that the exact solution for 
 the energies of the Hamiltonian (\ref{int}) 
 with
 $g={\eta}/{\sin(\alpha)}$   
 is 
 $$ E=2\sum_{j=1}^N v_j+gN\cos(\alpha), $$ 
 where the $\{v_j\}$ are solutions to (\ref{bae}). 
Making the change of variable
 $$\alpha\rightarrow \eta\alpha $$
 and taking the limit $\eta\rightarrow 0$ this reproduces the known
 exact solution \cite{richardson},
 given that
 $$\sum_{j=1}^\L b^\dagger_j b_j\phup  =N. $$
As a set of conserved operators for the Hamiltonian (\ref{int}) we 
take $\{t(\ve_k):\,k=1,\dots , \L\}$, which naturally 
generalise those of \cite{crs},
and arise in the solution of the inverse problem below. These operators
clearly satisfy 
$$[H_{RD},\,t(\ve_k)]=[t(\ve_l),\,t(\ve_k)]=0, 
~~~~\forall\,k,l=1,\dots , \L.$$ 
Hence the above construction shows that
the russian doll BCS model is both integrable and exactly
solvable. 

We note that the unitary time-reversal transformation of (\ref{int}) 
is equivalent to the 
change of variable $\alpha\rightarrow -\alpha$ with $g$
and each $\ve_j$ fixed.   
It is easily  seen  
that the change $\alpha\rightarrow -\alpha$ and $\eta\rightarrow
-\eta$ leaves  $g$, $E$ 
and each of the Bethe ansatz equations (\ref{bae}) invariant.  
We mention also that in the above the single particle energies
$\ve_j$ are entirely arbitrary. In particular, it is not necessary to
have $\ve_j<\ve_k$ for $j<k$. 

\section{Energy spectrum} 
To study the behaviour of the spectrum of (\ref{int})  
 as $\alpha$ varies between $0$ and $\pi/2$, we 
solved the Bethe ansatz equations (\ref{bae})  
for the picket fence model in 
which the energy levels are equally spaced, 
that is $\eps_j=j-(\L+1)/2$. With this convention the Fermi level is
zero for a system at half-filling.    The blocking effect simplifies
the calculation of the energy of the excited states,
since 
a state consisting of say $N'$ Cooper pairs and  two free electrons 
 blocking levels $m$ and $n$ has energy 
$$
E=2\sum_{j=1}^{N'} v_j+g N' \cos(\alpha)+\eps_m+\eps_n
$$
where the $\{v\}\equiv\{v_1,\dots,v_{N'}\}$ satisfy the following 
Bethe ansatz equations  
$$
\exp(-2i\alpha)\prod_{{{k=1}\atop{k \neq m,n}}}^\L
\frac{v_j-\ve_k-i\eta/2}{v_j-\ve_k+i\eta/2}
=\prod_{l\neq j}^{N'}\frac{v_j-v_l-i\eta}{v_j-v_l+i\eta}.
$$

For a fixed number of particles the excitations of the model
(\ref{int}) 
are classified according to the
initial distribution of the Cooper pairs and blocked energy levels at
$g=0$, thereby suggesting 
the initial configuration of 
the Bethe ansatz roots. 
For example, a solution of the Bethe ansatz equations for the ground state
at half-filling is found using the initial conditions 
 $v_j = \eps_j \ ,\
j=1,\dots,\L/2$, whereas an 
excited state formed by promoting a Cooper pair above the Fermi
level is obtained from the initial conditions 
$v_j=\eps_j \ ,\ j=1,\dots ,(\L{-}2)/2 $ and $
v_{\L/2}=\eps_{(\L+2)/2}$.  

We solved the Bethe ansatz
equations iteratively starting with each possible 
     configuration of the roots at $g=0$ and $\alpha=0$. 
Using this
method we calculated 
the excited state  spectrum for a model with  three Cooper pairs at
half-filling ($\L=6$), a total of 141 states if the particle-hole
symmetry 
is not taken into account. 
Figure \ref{Ealp} shows the energy levels as a function of the
 coupling $g$ for three different values of $\alpha$. 
 For small $\alpha$ and large coupling $g$, the energy levels form narrow,
well-separated bands in agreement with the results of Richardson's 
 model (e.g. see 
\cite{yba}). As
$\alpha$ increases the energy levels move, the original bands split
 apart, and new bands are formed.  Throughout, the  gap  
between the  ground state energy and the first group of excited
states remains, though it decreases as $\alpha$ increases. 
There is also a set of states for which the energy of each state does
 not depend on $\alpha$.
 These are the states that consist of one 
Cooper pair and all levels apart from levels 
$l$ and $m$ are blocked. 
The single Bethe ansatz
equation has two solutions:
$$
2v_{\pm} =  \eps_l+\eps_m \pm  \Bigl (g^2 + (\eps_l-\eps_m)^2\Bigr)^{1/2}
- g \cos (\alp) 
. 
$$
Consequently the  energy for this state is 
$$ 
E_{\pm}=\eps_l+\eps_m\pm  \Bigl (g^2 + (\eps_l-\eps_m)^2\Bigr)^{1/2}
+\sum_{i \in {\cal B} }\eps_i,
$$
where $ {\cal B}$ indexes the set of blocked energy levels. 
 
\begin{fig}[t]
\begin{center}
 \includegraphics{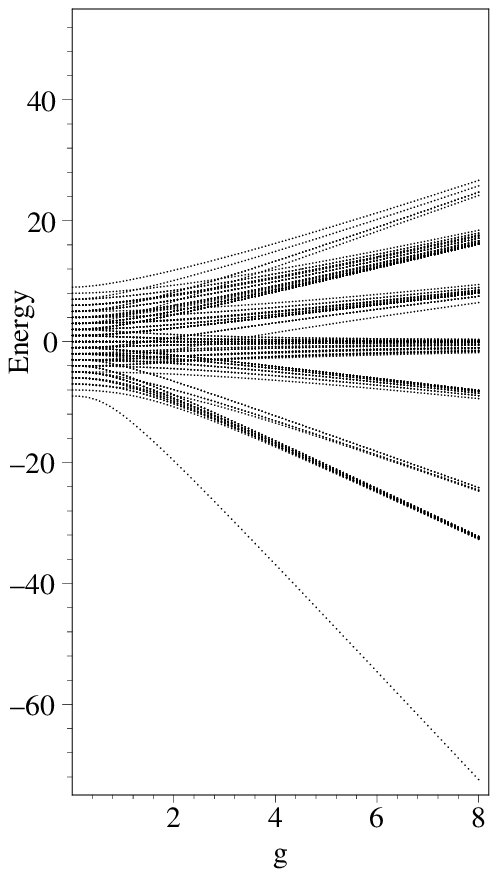}
 \includegraphics{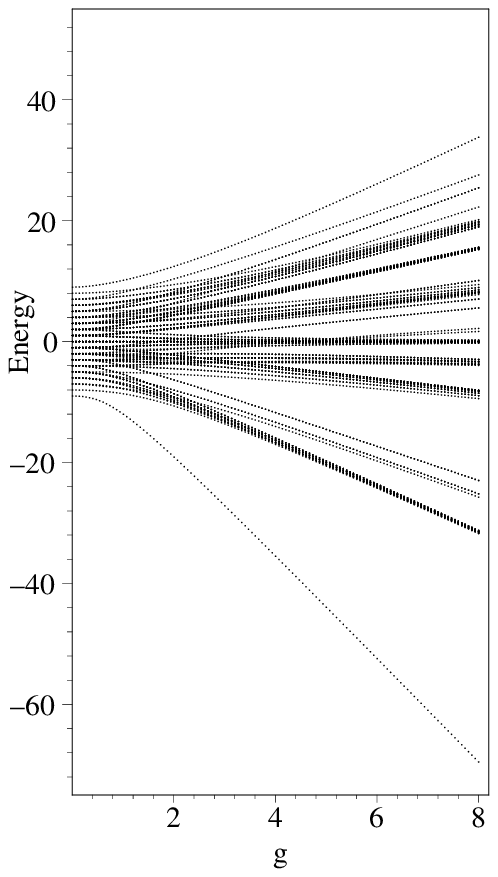}
 \includegraphics{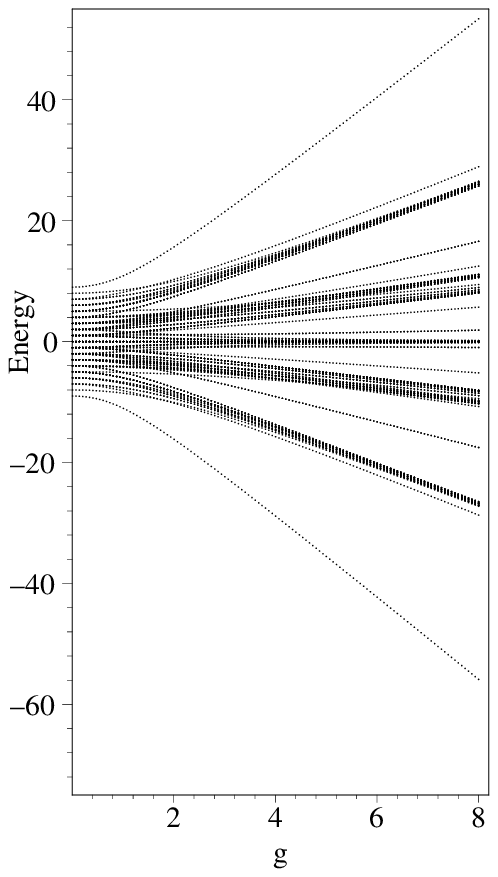}
\[
\begin{array}{ccc}
 \parbox{0.33333333\linewidth}{\quad \quad \quad 
   ~~~~~\protect\footnotesize\ref{Ealp}a:~$\alpha=~0.01$}& 
\parbox{0.33333333\linewidth}{\quad \quad \quad 
   ~\protect\footnotesize\ref{Ealp}b: 
$\alpha=~1.06$}&
\parbox{0.33333333\linewidth}{\quad \quad 
   \protect\footnotesize\ref{Ealp}c: 
$\alpha=~1.51$  }
\end{array}
\]
\caption{ \label{Ealp} The behaviour of the energy levels as $\alpha$
is increased. The results shown are for the picket fence model of $\L=6$
single particle energy levels at half-filling. As the 
coupling $g$ increases, the energy levels coalesce into
bands. Variations in  
$\alpha$ lead to different degeneracies of the bands. Note that in each case
the ground state is non-degenerate and there is no level crossing
between the ground and first excited energy level as $g$ increases. 
} 
\end{center}
\end{fig}

Despite the fact that 
the Bethe ansatz equations have to be solved
numerically to obtain the root positions  
for nonzero values of $g$, Gaudin \cite{gaudin3} and Sierra et
al.~\cite{sierra_exc} have made  
some conjectures as to the behaviour of the BCS roots at very large $g$
for Richardson's model.
Numerical studies indicate that 
as $g$ increases
some of the roots 
form complex conjugate pairs in a complicated pattern which may involve
 roots forming pairs, splitting apart, and forming new pairs, or
complex conjugate pairs separating and becoming independent real roots 
again.  The root behaviour  
of the ground state of a model with an even number of
Cooper pairs is very simple: all roots form complex conjugate pairs
and become infinite at large coupling.  For an odd number of
Cooper pairs the roots behave in the same way apart from the root initially
closest to the Fermi level, which remains finite.   
Generally most roots tend to infinity at
very large $g$ whether or not they have formed a complex conjugate
pair. The number of the remaining finite roots 
can be thought of as the number of elementary
excitations of the excited state. 
There is a formula due to Gaudin \cite{gaudin3}
that predicts the number of states with a particular number of finite
roots at $g = \infty$,  
and  an algorithm due to
Sierra et al. \cite{sierra_exc} that matches the initial configuration
of roots (or 
distribution of Cooper pairs over the available energy levels) 
to a final state.

It is clear from the splitting of the energy bands shown in  
Figure \ref{Ealp} that Gaudin's formula and the algorithm due to
Sierra et al. are not 
valid for nonzero $\alpha$.  In fact the russian doll BCS 
model exhibits the same general root behaviour as Richardson's 
model (as shown in Figures \ref{rt1}a-c) 
but many more roots tend to infinity as $g \to \infty$.  
In each of the figures the large $g$ configuration of Bethe ansatz roots
consists of two real roots and one complex conjugate pair of roots.  
For $\alpha=0.01$,   Figure~\ref{rt1}a
shows there are two finite real roots, in agreement  with 
the algorithm which predicts two elementary 
excitations  \cite{sierra_exc}. 
At $\alpha$ close to $\pi/2$ (see Figure \ref{rt1}c), it appears that a 
single real root remains finite, but upon close examination at $g\approx
100$, the root is seen to be very slowly diverging.  Thus there are no
longer any finite roots in this particular case.

\begin{fig}[t]
 \begin{center}
\[
\begin{array}{ccc}
 \parbox{0.33333333\linewidth}{\qquad \quad \quad \quad
   ~~~\protect\footnotesize\ref{rt1}a:~$\alpha=~0.01$}& 
\parbox{0.33333333\linewidth}{\qquad \quad \quad \quad 
   \protect\footnotesize\ref{rt1}b: 
$\alpha=~1.06$}&
\parbox{0.33333333\linewidth}{\qquad \quad \quad \protect\footnotesize\ref{rt1}c:
$\alpha=~1.51$  }
\end{array}
\]
\vspace{-0.9cm} 

  {\includegraphics{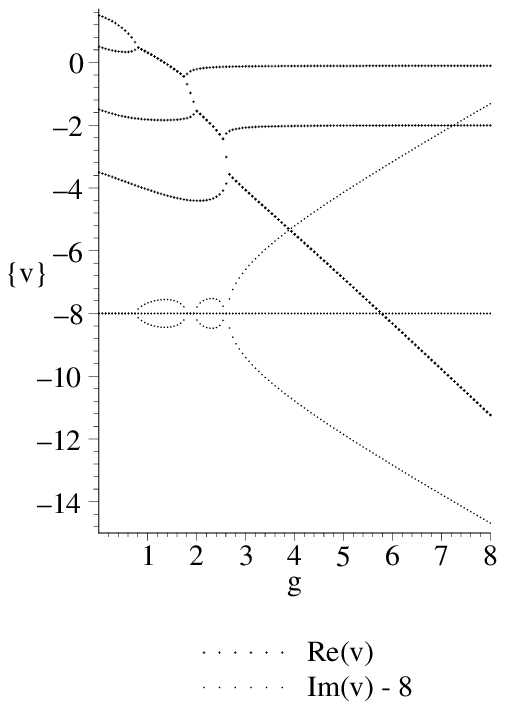}}%
  {\includegraphics{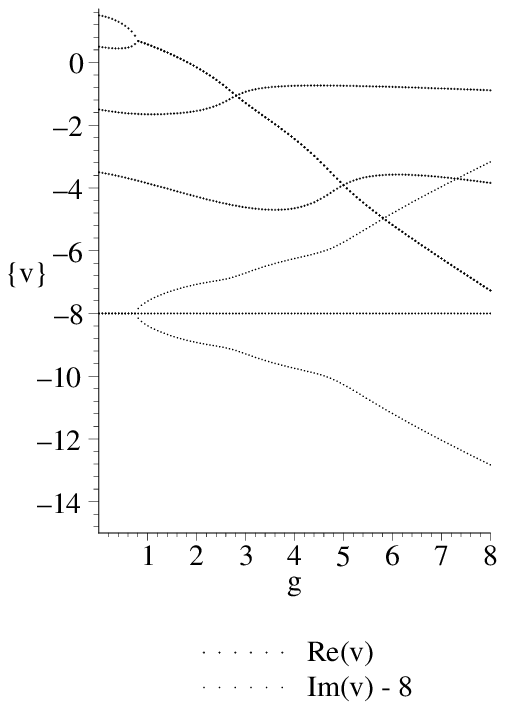}}%
  {\includegraphics{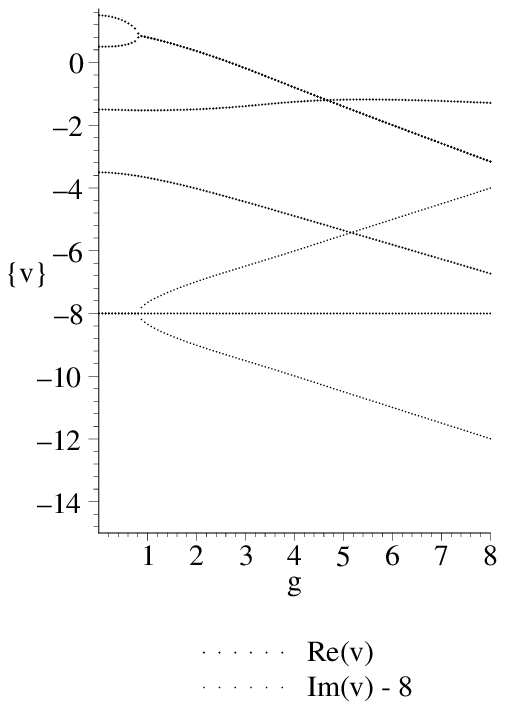}}%

\vspace{-0.4cm} 

\caption{ \label{rt1} The evolution of the real and imaginary parts of
  the roots  $\{v\}$ with increasing $g$. The three plots shown are
  for fixed values of $\alpha$.  In each case the roots correspond to 
  the excited state of the $\L=8$ model with 
$N=4$ Cooper pairs and initial conditions
$\{v\}=\{\eps_1,\eps_3,\eps_5,\eps_6\}$.}  
 \end{center}
 \end{fig}

\section{Solution of the inverse problem}   
In preparation for the next section we would like to be able to 
express any local operator in terms of the elements of the monodromy
matrix.  This is known as the inverse problem, and its solution 
will enable the form factors of the local operators to be determined. 
The results of this section are the generalisation of those in \cite{mt}
to the case where the operator $\g$ is included in the definition of the
monodromy and transfer matrices (cf. \cite{zlmg}). 

Hereafter we set $R_{jk}\equiv R_{jk}(\ve_j-\ve_k)$ 
for notational ease. 
First we note that
$$  T(\ve_j)
=  R_{j(j-1)}\cdots R_{j1}
P_{0j}\g_j R_{j\L}\cdots R_{j(j+1)} $$  
 or equivalently
 \eq P_{0j}= 
 R_{j1}^{-1}\cdots R_{j(j-1)}^{-1} T(\ve_j) R_{j(j+1)}^{-1} \cdots
 R_{j\L}^{-1} 
 \g^{-1}_j \label{p0j}. \en
It is readily deduced that   
   \eq t(\ve_j)
        =  R_{j(j-1)}\cdots R_{j1}
              \g_j R_{j\L}\cdots R_{j(j+1)}. \label{tau} \en
At this point we appeal to the unitarity condition
(\ref{unitarity}) 
which allows us to write
$$t(\ve_j)=
              R^{-1}_{(j-1)j}\cdots R^{-1}_{1j} \g_j
              R_{j\L}\cdots R_{j(j+1)}. $$
Defining
$$\Gamma_{jk}=R_{1k}R_{2k}\cdots R_{jk},  $$
we then have 
$$t(\ve_j)= 
\Gamma^{-1}_{(j-1)j} \g_j R_{j\L}\cdots R_{j(j+1)}, $$
which leads to the formula  
\beq \prod_{j=1}^k t(\ve_j)= \Biggl(\prod_{j=1}^k
\g_j\Biggr)
\Gamma_{k\L}\Gamma_{k(\L-1)}\cdots\Gamma_{k(k+1)}.
\label{prod} \eeq  
Equation (\ref{prod}) is proved by induction, noting first that 
\bea t(\ve_1)&=& \g_1 R_{1\L}R_{1(\L-1)}\cdots R_{12} \nn
               \\
               &=& \g_1 \Gamma_{1\L
               }\Gamma_{1(\L-1)}\cdots \Gamma_{12}. \nn \eea 
Employing equation (\ref{prod})  we find
\eq \Biggl(\prod_{j=1}^{k-1}t^{-1}(\ve_j)\Biggr) P_{0k}
\Biggl(\prod_{j=1}^{k-1} t(\ve_j)
\Biggr ) 
  = \Gamma^{-1}_{(k-1)k} P_{0k} \Gamma_{(k-1)k}. \label{gamp}  \en
Combining (\ref{p0j}) and (\ref{tau}) we have
$$  
P_{0k}=  R_{k1}^{-1}\cdots R_{k(k-1)}^{-1} T(\ve_k) t^{-1}(\ve_k)
   R_{k(k-1)} \cdots R_{k1}  
$$
from which  we obtain the following expression
$$
    T(\ve_k) t^{-1}(\ve_k)
    = \Gamma^{-1}_{(k-1)k} P_{0k} \Gamma_{(k-1)k}. \nn
     $$  
      Thus using (\ref{gamp})  we have
      \eq P_{0k}=
      \Biggl(\prod_{j=1}^{k-1} t(\ve_j)\Biggr) T(\ve_k)\Biggl(\prod_{j=1}^k
      t^{-1}(\ve_j)\Biggr).  \label{p0k} \en
      Finally, since
$$
P_{0k}=
{\renewcommand{\arraystretch}{1.2}
\left ( \begin{array}{cc}
N_k\phup & b_k\phup  \cr  b_k^\dagger & I-N_k\phup  
\end{array}
\right)
}
$$ 
  formula (\ref{p0k}) leads to expressions for the local operators
$b_k,\,b_k^\dagger,\,N_k$
      terms of
      the global operators $A(u),\,B(u),\,C(u)$ and $D(u)$
      of the algebraic Bethe ansatz.

\section{Pair occupation fluctuations}

\label{corr}
In this section 
we use the notation $\langle \chi \rangle =
 \langle \{v\} | \chi | \{ v\} \rangle \,/\,  \langle \{ v\} |\{ v\}
\rangle$ for
 any operator $\chi$ where
 $| \{v\} \rangle$ is a Bethe state associated with the roots 
 $ \{v\}\equiv\{v_1,\dots,v_N\}$.
For the BCS model 
it is of interest to study the behaviour of the correlation function 
$ C_k$ describing the fluctuations in the Cooper pair occupation numbers for
each level $k$: 
\bea
C_k^2 &=& \langle N_k^2\rangle -\langle N_k\rangle ^2 \nn \\
&=& \langle I-N_k \rangle  \langle N_k \rangle  
. \nn  \eea 
For Richardson's model ($\alp=0$), this
correlation function has been studied in \cite{vdr,ao}.

Solving the inverse scattering problem
allows the correlator to be expressed in terms of the form factor of
the global operator $D(u)$ via
$$
\langle \{ v\} | 1{-}N_k | \{ v\} \rangle  =
 \langle \{ v\}| D(\ve_k) t^{-1}(\ve_k)|  \{ v\} \rangle.  
$$ 
Fortunately the Slavnov formula \cite{slavnov} (see also
\cite{korepin,kmt,lzmg}) for the scalar products of
Bethe  states  
leads to explicit determinant representation for the form factors
of the operators $A(u),\,B(u),\,C(u)$ and $D(u)$.  
 We follow \cite{lzmg}, in which the Slavnov formula is 
described for $\t(u-i\eta/2)$ together with form factors for the
elements of the 
associated monodromy matrix (up to a rescaling of $\eta$). 
We  define it here in terms of   
the shifted functions 
$$a(u)=\tilde{a}(u-i\eta/2),~~~~~~~d(u)=\tilde{d}(u-i\eta/2) $$ 
where $\tilde{a}(u)$ and $\tilde{d}(u)$ are given by (\ref{ta}) and
(\ref{td}) respectively. 

Below both sets of parameters $\{w_i\}$ and $\{v_i\}$ are assumed 
to be solutions of the Bethe ansatz equations (\ref{bae}).  
The scalar products of the Bethe states are obtained from 
$$
\langle\{ w\}   |\{ v\}\rangle=
\frac {\det F}{\prod^N_{k>l}(v_k-v_l)
\prod^N_{i<j}(w_i-w_j)},
$$
where $F$ is an $N \times N$ matrix with entries
$$ 
F_{ij}=\frac{i\eta d(w_i)}{(v_j-w_i)}\biggl( a(v_j)\prod_{k\neq
i}^N (v_j-w_k+i\eta)-  d(v_j)\prod^N_{k\neq i}(v_j-w_k-i\eta)\biggr),
$$ 
unless $\{ w\} = \{ v\}$, in which case the diagonal entries become
\bea 
F_{ii}&=&d(v_i) \Biggl (\sum^N_{l\neq i}
\frac{ a(v_i)}{v_i-v_l+i\eta}\prod_{k=1}^N(v_i-v_k+i\eta)
+\sum_{l\neq i}^N
\frac{
d(v_i)}{v_i-v_l-i\eta}\prod_{k=1}^N(v_i-v_k-i\eta)\Biggr)~
 \nn \\
 &&~~~~+d(v_i) \Biggl (a'(v_i)\prod_{k=1}^N(v_i-v_k+i\eta) 
+ d'(v_i) 
\prod_{k=1}^N(v_i-v_k-i\eta) \Biggr ) ,
\label{diag}  \eea  
where the prime denotes the derivative. 
Note that (\ref{diag}) corrects typographical errors appearing in
equation (46) of \cite{lzmg}. 
The matrix elements of the operator $\tilde{D}(u)$ were derived in  
\cite{lzmg} and read 
$$
 \fl
\langle\{ w\}|\tilde{D}(u)|\{ v\} \rangle
=
\frac{\Theta(u)  \tilde{d}(u)}{\prod^N_{k>l}(v_k-v_l)   \prod^N_{i<j}(w_i-w_j)}
\l(\prod_{i=1}^N\frac{u-w_i-i\eta/2}{u-w_i+i\eta/2}\r)
\det \l(F+Q(u)\r)  
$$
where
$$
\Theta(u)=\prod_{k=1}^N \frac{u-w_k+i\eta/2}{u-v_k+i\eta/2}
$$
and 
$$
 Q_{ij}(u)=\frac{ i\eta  d(w_i)  d(v_j)}{
(u-w_i+i\eta/2)(u-w_i-i\eta/2)} 
 \Biggl [ 1-\frac{\tilde{a}(u)}{\tilde{d}(u)}  \prod_{k \neq i}^N
\frac{u-w_k+3i\eta/2} {u-w_k-i\eta/2}\Biggr]  \prod_{l=1}^N (v_j-v_l -i\eta).
$$
In the limit $u\to \eps_k$ the term 
within the square brackets becomes equal to unity  since
$\tilde{a}(\ve_k)=0$, and   
 the  expression for the form
factor of the local operator $N_k$ between identical 
Bethe eigenstates is also very simple: 
$$ 
\langle N_k \rangle 
=    1- \frac{\det
(F+Q(\eps_k))}{\det F}.  
$$

Figure \ref{corrp} shows the behaviour of $C_k$ for the ground state 
of the $\L=6$ system  
at half-filling, for three values of the coupling
strength $g$ as the parameter
$\alpha$ is increased from just above  zero to just below $\pi/2$.   
\begin{fig}[t]
 \begin{center}
\resizebox{.75\textwidth}{!}%
{  {\includegraphics{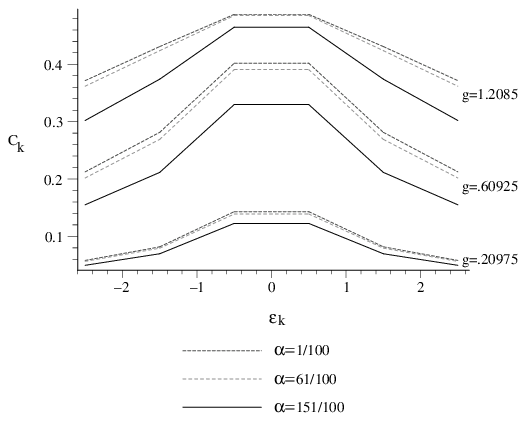}}}

\vspace{0.2cm}

\caption{ \label{corrp} The  correlation function 
$C_k$ for various  choices of 
$\alpha$ and the coupling strength $g$.
The results shown are for the ground state of the $\L=6$ model at half-filling. 
In each case the distribution of $C_k$ peaks about the Fermi 
level. For fixed $g$, increasing values of $\alpha$ suppress the
magnitude of $C_k$ across all single particle energies $\ve_k$.} 
 \end{center}
 \end{fig} 

\section{Conclusion} 

We have demonstrated the integrability of the russian doll BCS model
\cite{lrs},  which generalises Richardson's model by the inclusion of phases in
 the pair scattering 
couplings, parameterised by a variable $\alpha$. 
We have applied the algebraic Bethe ansatz to find the exact
solution of the model, and have also solved the inverse problem for the
computation of form factors and correlation functions. In particular,
an explicit expression for the one-point correlator characterising Cooper pair
occupation fluctuations was obtained.  Analysis of this result
indicates that the ground state structure is qualitatively independent
of $\alpha$, and increasing $\alpha$ suppresses the fluctuations
across all single particle energy levels. On the other hand, our
analysis of the energy spectrum shows that the degeneracies of the
energy bands in the strong coupling limit is dependent on $\alpha$. An
open problem is to determine the extent to which $\alpha$ affects 
the thermodynamic properties of the model.  
The quantum transfer matrix method and the solution of the associated
 nonlinear integral 
equations constitute powerful techniques  
for exactly calculating
the free energy of many integrable systems.  
These techniques  have successfully been applied to the cases of the 
XXZ model
\cite{st} and an integrable spin ladder \cite{bgostf}. 
As we have shown that the Hamiltonian is embedded in the transfer matrix
as the second-order term in the series expansion about infinite spectral
parameter, we hope that our results are a first step in formulating
the quantum transfer matrix method for 
the russian doll BCS model.   

\vspace{0.5cm}
\noindent{\bf Acknowledgements --} Financial support from the
Australian Research Council is gratefully acknowledged.   
We thank David De Wit, Germ\'an Sierra
and Huan-Qiang Zhou for helpful comments.

\end{document}